\documentstyle[prl,twocolumn,aps,epsf]{revtex}
\newcommand{\be}{\begin{equation}}
\newcommand{\ee}{\end{equation}}
\newcommand{\ep}{\epsilon}
\newcommand{\r}{{\bf{r}}}
\begin{document}

\twocolumn[\hsize\textwidth\columnwidth\hsize\csname @twocolumnfalse\endcsname
\draft
\tolerance 5000

\title{Landau diamagnetism and magnetization of interacting diffusive conductors}

\author{Gilles Montambaux}

\address{Laboratoire de Physique des Solides, Associ\'e au CNRS, Universit\'e
Paris Sud,  91405 Orsay, France}

\maketitle

\begin{abstract} {
We show how the orbital magnetization of an interacting disordered diffusive electron gas can be simply related to the magnetization of the non-interacting system having the same geometry. This result is applied to the persistent current of a mesoscopic ring and to
the relation between Landau diamagnetism and the interaction correction to the magnetization of diffusive systems. The
field dependence of this interaction contribution can be deduced directly from the de Haas-van Alphen oscillations of the
free electron gas. Known results for the free orbital magnetism of finite systems can be used to derive the interaction
contribution in the diffusive regime in various geometries.}
\end{abstract}

\pacs{PACS Numbers: }
]

In recent years, there has been many theoretical works on the
thermodynamic properties of mesoscopic electronic systems, in
particular concerning their orbital magnetism
\cite{Aslamasov74,Altshuler83,Oh91}. The simplest description of
metals deals with non-interacting electrons in the absence of
disorder. The correction to Landau susceptibility due to
electron-electron interactions and phase coherence has been worked
out by Altshuler {\it et al.}\cite{Altshuler83}. Similarly, the
persistent current in mesoscopic rings has been extensively
studied. The simplest description of this effect was first done in
a strictly one-dimensional (1D) picture of non-interacting
electrons\cite{Cheung88} and the effect of diffusion and
interaction was described later by Ambegaokar and
Eckern\cite{Ambegaokar90} and Schmid\cite{Schmid91}.

The very simple approach for free electrons and the more
sophisticated description of interacting electrons in a disordered
potential have been developed in a completely independent way.
Here, we show how these descriptions are closely related. The main
result of this letter is a simple relation between the response of
a clean {\it non-interacting} electron gas and the response of a
{\it diffusive} electron system {\it in the presence of
interactions}. This result originates from the very similar
structures of the Schr\"odinger equation and of the diffusion
equation which describe the two systems.

As a first example, we show how the persistent current of a 1D
ballistic ring is related to the current of a quasi-1D diffusive
ring in the presence of interactions\cite{q1D}. Then we show how
the interaction contribution to the orbital magnetism of any
diffusive system can be deduced immediately from the orbital
response of the same non-interacting system. As a second example,
we show how the interaction contribution to the susceptibility of a
 bulk diffusive system is derived directly from the Landau susceptibility.
Then, from the de Haas-van Alphen oscillations of the free electron
gas, we deduce the field dependence of the interaction induced
magnetization. Finally we use this mapping to derive the finite
size corrections (in $L_\varphi
/L$) in the diffusive case from the $1/k_F L$ corrections of the
magnetization of the clean system.
\bigskip

Classically, the probability $p(\r,\r',\omega)$ for a particle to diffuse from a point $\r$ to another point $\r'$
is solution of the diffusion equation
\be
(-i \omega+\gamma-D \nabla_{\r'}^2) p_\gamma(\r,\r',\omega)=\delta(\r-\r')
\label{diffusion}
\ee
$D$ is the diffusion coefficient. This probability has actually two parts, a purely classical one (the Diffuson) and an
interference part (the Cooperon) which results from interference between time reversed trajectories. The Cooperon  has  to
been taken at $\r=\r'$. In a magnetic field, it obeys eq.(\ref{diffusion}) where $\nabla$ has to be replaced by $\nabla + 2
i e {\bf A}/\hbar c$, ${\bf A}$ being the vector potential. The charge $(-2e)$ accounts for the pairing of time reversed
trajectories which are supposed to propagate coherently up to a time $\tau_\phi$. $\gamma=1/\tau_\phi$ and  $L_\varphi
=\sqrt{ D \tau_\varphi}$ is the phase coherence length.

The probability $p(\r,\r',\omega)$  has the same structure as the disordered averaged  (retarded) Green's function
$\overline G^R_\epsilon(\r,\r',\ep)$ of
the Schr\"odinger
equation for a free particle of energy $\ep$ and charge $-e$ in a disordered potential:
\be
(\ep-i{  \hbar \over 2 \tau_e}+{\hbar^2 \over 2 m} \nabla_{\r'}^2) \overline G^R_\ep(\r,\r')=\delta(\r-\r')
\label{green}
\ee
where $\tau_e$ is the elastic mean free path. In a field, $\nabla \rightarrow \nabla +  i  e{\bf A}/\hbar c$.
The solutions of eqs.(\ref{diffusion},\ref{green}) can be written as
\be
p_\gamma(\r,\r')=p_\gamma(\r,\r',\omega=0)=\sum_n {\psi^*_n(\r)\psi_n(\r') \over \gamma +E_n^d}
\ee
and

\be
\overline G^R(\r',\r,\ep)=\sum_n {\psi^*_n(\r)\psi_n(\r') \over \ep+i{  \hbar \over 2 \tau_e}-E_n^s}
\ee
where the eigenvalues $E_n^{d,s}$ are the solutions of similar equations

\be -D \Delta \psi_n= E_n^d \psi_n \ \ \ , \ \ -{\hbar^2 \over 2 m} \Delta \psi_n=E_n^s \psi_n
\ee
with the mapping from the diffusion to the Schr\"odinger problem:
\begin{eqnarray}
D &\rightarrow& {\hbar \over 2 m} \nonumber \\
2 e &\rightarrow& e \nonumber \\
\hbar \gamma &\rightarrow& -\ep -i{\hbar \over 2 \tau_e}
\label{subst}
\end{eqnarray}
It has long been recognized that a Diffuson (or a Cooperon) behaves like a free particle with an effective mass
$m^*=\hbar/2D$. The goal of this letter is to study the consequences of this mapping on the orbital magnetism of clean and
diffusive systems.

For a disordered finite system of size $L$, the Thouless energy
$E_c$, given by $\hbar D /L^2$, is equivalent to the mean
interlevel spacing $\Delta=\hbar^2/2 m L^2$ of the eigenvalues of
the Schr\"odinger equation. More interesting is the relation
deduced from eq. (\ref{subst}).

\be
{L \over L_\varphi} \rightarrow - i k_F L -{L \over 2 l_e}
\label{L12}
\ee
where $l_e=v_F \tau_e$. $1/\tau_e$ spreads the levels of the Schr\"odinger equation while $1/\tau_\varphi$ spreads those of
the diffusion equation.
Inelastic disorder on the Cooperon plays thus the same role as elastic disorder on a free particle. More important, the
relation (\ref{L12}) expresses that the limit $k_F L \gg 1$ for the clean system corresponds to the macroscopic limit $L \gg
L_\varphi$.
Inversely the mesoscopic limit $L \ll L_\varphi$ corresponds to having only one Schr\"odinger particle  in a box ($k_F L \ll 1$).

Let us now apply this mapping to the calculation of the
magnetization. First, the $T=0K$ magnetic moment of the free
electron gas (including spin) can be written as \be M={\partial
\over
\partial B } \cal{N}(\ep_F,B)
\label{M11}
\ee
where $\cal{N}(\ep,B)$ is the double integral of the total density
of states $\rho(\ep,B)$. This contribution is known as the Landau
magnetization. Then taking into account electron-electron
interactions in the Hartree-Fock picture gives an additional
contribution\cite{Aslamasov74}. For a completely screened
interaction $U(\r-\r')=U \delta(\r-\r')$ \cite{screening}, this
contribution is given by

\begin{eqnarray}
\langle M_{ee} \rangle &=& - { U \over 4} {\partial \over \partial B} \int \langle n^2(\r) \rangle d\r \nonumber
\\
&=& - U {\partial \over \partial B} \int \langle \rho(\r,\omega_1)\rho(\r,\omega_2)\rangle d\r d\omega_1 d\omega_2
\end{eqnarray}
This expression contains the Hartree and Fock contributions. $n(r)$ is the local density. $\rho(\r,\omega)$ is the local density of states (per spin direction). The average product $\langle \rho(\r,\omega_1)\rho(\r,\omega_2)\rangle$ is nothing but the Fourier transform of the return probability $p_\gamma(\r,\r,t)$, so that one gets finally \cite{Montambaux95}:

\be
\langle M_{ee} \rangle(\gamma) =-{\lambda_0 \hbar \over \pi}{\partial \over \partial B} \int {P_\gamma(t) \over t^2} dt
\label{Mee}
\ee
where $P_\gamma(t)=\int p_\gamma(\r,\r,t) d\r$ in the space integrated return probability. $\lambda_0=U \rho_0$ is a dimensionless
interaction parameter and $\rho_0$ is the
average density of states (per spin direction).
Writing the density of states as
\be
\rho(\ep)=-{1 \over \pi} \int \mbox{Im} \overline G^R\ep(\r,\r) d\r
\ee
and the integrated return probability as
\be
\int P_\gamma(t) dt = \int p_\gamma(\r,\r) d\r
\ee
one obtains immediately from eqs.(\ref{M11},\ref{Mee}) that the two magnetizations are related
(since the $1/t^2$ term in eq.(\ref{Mee}) is equivalent to a double integral over $\gamma$)

\begin{equation}
M \, \tilde{=} -  {1 \over \lambda_0} \mbox{Im}
[\langle M_{ee} \rangle (\gamma=-{\epsilon_F \over  \hbar}-i0) ]
\label{mapping}
\end{equation}
The sign $\tilde{=}$ means that the two quantities are equal, provided the substitutions (\ref{subst}) have been made.
It should then be remembered   that eq.(\ref{Mee})
 corresponds to taking the first order contribution in $\lambda_0$ to the grand potential.
It is known that taking into account higher diagrams in the Cooper channel, one has to renormalize the interaction
parameter which becomes energy dependent $\lambda(\ep)$\cite{Altshuler85b,Eckern91,repulsive}:

\be
\lambda_0 \rightarrow \lambda(\ep)=\lambda_0/(1+ \lambda_0 \ln {\ep_F \over  \ep})=1/\ln {T_0 \over \ep}
\ee
where $T_0$ is defined as $T_0=\ep_F e^{1/\lambda_0}$. Then the relation (\ref{mapping}) can be simply modified as:

\begin{equation}
M = - \mbox{lim}_{\lambda_0 \rightarrow 0} {1 \over \lambda_0} \mbox{Im}
[\langle M_{ee} \rangle (\gamma=-{\epsilon_F  \over \hbar}-i0) ]
\label{mapping2}
\end{equation}

\bigskip

As an example, we consider the case of a 1D diffusive ring of perimeter $L$ pierced by a Aharonov-Bohm flux $\phi$. Starting from the flux dependent part of the
return probability

\be
P(t)={L \over 4 \pi D t} \sum_{p=1}^\infty e^{-{p^2 L^2\over 4 D t}}\cos 4\pi p \varphi
\ee
where $\varphi=\phi/\phi_0$, $\phi_0$ being the flux quantum, one simply gets from eq.(\ref{Mee}), the harmonic dependence
of the average persistent current due to interactions.

\be
\langle I_{ee} \rangle=16 \lambda_0 {E_c \over \phi_0}\sum_{p=1}^\infty {1 \over p^2} (1+p{L \over L_\varphi})e^{-p L
/L_\varphi}\sin 4 \pi p \varphi
\ee
This result, for $L_\varphi=\infty$, was first obtained by Ambegaokar and Eckern (AE) \cite{Ambegaokar90}. It was then
generalized to the case where $L_\varphi$ is finite\cite{Montambaux95}. Using the relation (\ref{mapping}), one deduces
immediately the average persistent current for a clean 1D ring (clean means here that there is no diffusion. Disorder is
only
taken into account by a finite mean free path $l_e=v_F \tau_e$):

\be
 I = {2 \over \pi} I_0 \sum_{p=1}^\infty {1 \over p} (\cos p k_F L - {\sin p k_F L \over p k_F L})e^{-p L/2l_e} \sin 2 \pi p
\varphi
\ee
with $I_0= e v_F /L$.
This result  has first been obtained for the case $k_F L\gg 1$ (in this case, the $\sin x/x$ term cancels) in the absence of
disorder ($l_e=\infty$)\cite{Cheung88}. Note that the correspondance between the AE current and the current of the ballistic
ring is not  trivial.  The leading term in $k_F L$ for the clean case originates from the leading term in $L/L_\varphi$ in
the diffusive case. Therefore, taking simply the AE result for the mesoscopic limit $(L_\varphi=\infty$) would not have
produced the correct result for the clean ring. In other words, the $k_F L \gg 1$ limit corresponds to the macroscopic
limit for the diffusive case.
We will return to this point later where we will show how to derive the $L_\varphi/L$ corrections to diffusive
magnetization from perimeter corrections in the ballistic case.

\bigskip

Deducing the magnetization of a clean system from the one of the interacting system may not appear as the most useful
procedure. More interesting is to deduce the properties of an
interacting medium from those of the non-interacting one, {\it i.e.} to invert eq.(\ref{mapping}). This inversion is given
by:

\be
\langle M_{ee} \rangle \ \tilde{=} -{\lambda_0 \over \pi} \int_0^\infty  {M(\ep) \over \ep + \hbar\gamma} d\ep
\label{map1}\ee
with the substitution (\ref{subst}).
Defining  $\tilde{M}$ the magnetization of a free particule of mass $\hbar/2D$ and charge $2e$, so that $\tilde{M}(\ep) \,
\tilde{=}\,  M(\ep)$, one can rewrite

\be
\langle M_{ee} \rangle  = -{\lambda_0 \over \pi} \int_0^\infty  {\tilde{M}(\ep) \over \ep + \hbar\gamma} d\ep
\label{map2}\ee
Again, recognizing that
Cooper Channel renormalization modifies the interaction parameter,
  the energy dependence of this parameter can be incorporated exactly in the integral so that:
\be
\langle M_{ee} \rangle  = -{1 \over \pi} \int_0^\infty \lambda(\ep) \, {\tilde{M}(\ep) \over \ep + \hbar\gamma}\, d\ep
\label{central}
\ee
\bigskip

This is the main result of this paper. It gives straightforwardly {\it the magnetization of an interacting electron gas in
terms of the magnetization of the same non-interacting system}.

As a example, we now consider the orbital response of a 2D clean system. The (spinless) Landau susceptibility gives the non
oscillating part of the orbital response. It is given by
$\chi(\ep) = -e^2/(24\pi  m)$ and is independent of the energy $\chi(\ep)=\chi_L$.
Then, using the mapping (\ref{subst}), the susceptibility of the cooperon is $\tilde{\chi}(\ep)=-{4 \pi \over 3} {\hbar D
\over \phi_0^2}= - 4 \, \chi_L (\ep_F \tau_e)/\hbar$. From eq.(\ref{central}), one immediately deduces the interaction part of
the
susceptibility\cite{Altshuler83,Oh91}:

\begin{equation}
\chi_{ee} = \frac{4}{3} \frac{\hbar D}{\phi_0^2} \ln {\ln T_0 \tau_\varphi/\hbar  \over \ln T_0 \tau_e/\hbar}=
4 |\chi_L| {\ep_F \tau_e \over \hbar} \ln {\ln T_0 \tau_\varphi/\hbar  \over \ln T_0 \tau_e/\hbar}
\end{equation}
An ultraviolet cut-off $1/\tau_e$ has been added in order to cure the divergence at large energy.

In 3D, the Landau susceptibility becomes energy dependent
$\chi(\ep) = -e^2 k_F(\ep)/(24\pi^2  m)\propto \sqrt{\ep}$, so is the susceptibility $\tilde \chi(\ep)$ of the cooperon,
$\tilde \chi(\ep)= - 8 \chi_L \sqrt{\ep \tau_e \over 3 \hbar}$. Contrary to the $2D$ case where the susceptibility was
constant in
energy and of order $\ep_F \tau_e$, integration in energy gives here a much smaller contribution. Using eq.(\ref{central}),
one gets the interaction correction
in 3D:
\be
{\chi_{ee} \over  |\chi_L|}={16 \over \pi \sqrt{3}}{1 \over \ln T_0 \tau_e/\hbar}
 \ee
Consider again the 2D clean case. In addition to the Landau contribution, the de Haas-van Alphen effect expresses the
oscillatory behavior of the grand potential in $1/B$, with the fundamental period $1/B_0=e\hbar/m\ep_F$.  The grand
potential is given by\cite{Revue}:
\be
\delta {\cal A}(B) =-{1\over 2} \chi_L B^2  \left( 1 + {12 \over \pi^2} \sum_{s=1}^\infty {(-1)^s \over s^2}   \cos {2 \pi
s \ep_F
\over \hbar \omega_c} \right)
\ee
and the magnetic moment at fixed Fermi energy is given by $M=-\partial \delta {\cal A} / \partial B$. Its dependence versus field has the well-known saw-toothed behavior. One may wonder how this behavior translates into
the language of interacting diffusive electrons.
To simplify, we restrict ourselves to the first order in $\lambda_0$. Using the mapping (\ref{map1},\ref{map2}), one
deduces the interaction contribution to the magnetization,
 in units of $\lambda_0 \hbar D /\phi_0^2$:
\be \langle M_{ee} \rangle = {4 \over 3 } B \ln {\tau_\varphi \over \tau_e} + {8  \over \pi^2}{\partial \over \partial B}
B^2 \sum_{s=1}^\infty {(-1)^s \over s^2} f(2 \pi s{B_\varphi \over B})
\label{ALDHVA}
\ee
where $f(x)=-\mbox{ci}(x) \cos( x) -\mbox{si}( x) \sin (x)$.  The fundamental frequency $B_0$ has been transformed into the
characteristic field $B_\varphi=\hbar/(4 e D \tau_\varphi)$.
Alternatively, this magnetization could have been obtained from eq.(\ref{Mee}), with the following expansion of the return
probability in a constant field:

\begin{eqnarray}
P(t)&=&{B/\phi_0 \over \sinh 4 \pi B D t / \phi_0} \nonumber \\
&=&{1 \over 4 \pi D t}\left(1 + 2 \sum_1^\infty (-1)^s {t^2 \over t^2 + a^2 s^2} \right)
\end{eqnarray}
with
$a=\phi_0 / (4 BD )$. The result (\ref{ALDHVA}) can be easily generalized to all orders in $\lambda_0$ by considering the explicit dependence $\lambda(\ep)$ is eq.(\ref{central}).

\begin{figure}[!ht]
\centerline{
\epsfxsize 7cm
\epsffile{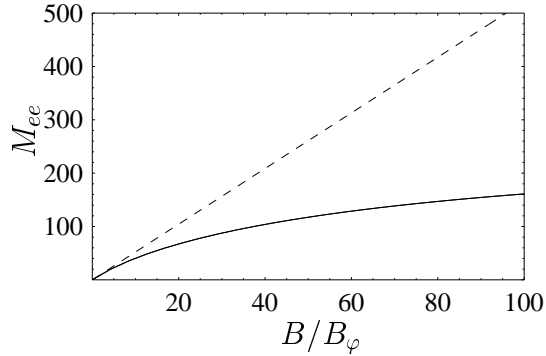}}
\medskip

\caption{Magnetization of  a diffusive interacting electron gas calculated to first order in $\lambda_0$, in units of
$\lambda_0 \hbar D/\phi_0^2$. The dashed line shows the linear low field behavior,   see eq. (\protect{\ref{ALDHVA}}).         } \label{AL0}
\end{figure}
\bigskip
\begin{figure}[!ht]
\centerline{
\epsfxsize 7cm
\epsffile{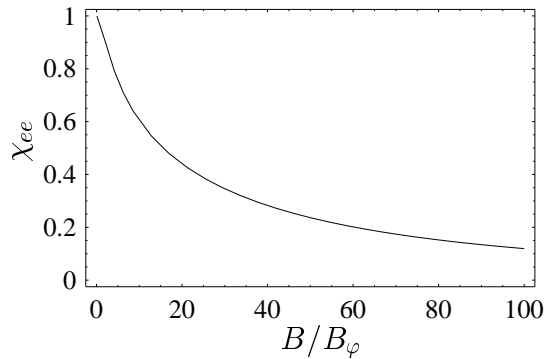}}
\medskip

\caption{Susceptibility of a diffusive interacting electron gas in units of $\lambda_0 \hbar D /\phi_0^2$. The amplitude at
zero field is $4/3 \ln \tau_\varphi /\tau_e$, see eq. (\protect{\ref{ALDHVA}}).}
\label{AL}
\end{figure}
\bigskip

Let us finally note that the limit $k_F L \gg 1$ for the
Sch\"odinger equation corresponds to the macroscopic regime $L \gg
L_\varphi$ for the diffusion equation. The opposite, so called
mesoscopic regime $L \ll L_\varphi$ would correspond to $k_F L \ll
1$, for which only the ground state is occupied. In the diffusive
context, this ground state is called the {\it zero mode}. The
cross-over between the mesoscopic regime where only a few modes are
relevant to the macroscopic regime where there is a quasi-continuum
of diffusion modes is quite difficult to describe\cite{AGIM}. It is
then quite useful to know the finite size $1/k_F L$ corrections to
the Landau susceptibility which have been extensively studied
\cite{vanleeuwen}. These corrections are usually of the form
\cite{vanleeuwen}:
\be \chi(L) \simeq \chi(\infty)\left(1 - {\alpha \over k_F L}\right)
\ee
Thus, knowing the finite size corrections to the Landau diamagnetism, one can  get the  $L_\varphi/L$ corrections to the bulk
susceptibility $\chi_{ee}$. For $L \gg L_\varphi$, they are of the form \cite{Boundaries}:
\be \chi_{ee}(L) \simeq \chi_{ee}(\infty)\left(1 - \alpha {L_\varphi\over L}\right)
\ee

In conclusion, we have shown that the magnetization of a diffusive
interacting electron gas can be deduced from the magnetization of
the non-interacting system. This mapping allows the study of finite
size properties of diffusive systems, in particular the cross-over
between the macroscopic and the mesoscopic regimes.


\begin{thebibliography}{99}
\bibitem{q1D} The diffusive ring
is quasi-1D in the sense that it is described by a multichannel 3D
Schr\"odinger equation with a random potential, but the equation
which describes the diffusive motion along the ring is
one-dimensional.
\bibitem{screening}We assume perfectly screened interaction for simplicity. Conclusions are not affected by this
hypothesis.
More generally, $U$ is a function of $k_F/\kappa$ where $\kappa$ is
the Thomas-Fermi vector\cite{Ambegaokar90}.
\bibitem{repulsive} We limit ourselves to the case of repulsive interactions, $\lambda_0 >0$.
\bibitem{Boundaries} The boundary conditions are different: for the clean isolated system trapped
within infinite well, the
Schr\"odinger equation  assumes Dirichlet boundary conditions while  the diffusion assumes the
condition of zero current.
\end{thebibliography}
\end{document}